\documentclass[conference]{IEEEtran}
\input epsf
\usepackage{graphicx}
\graphicspath{{figures/}}
\usepackage{hyperref}
\usepackage{graphicx}
\usepackage{amssymb}
\usepackage{amsmath}
\usepackage[super]{nth}

\usepackage{xspace}

\usepackage{color}
\definecolor{fgred}{rgb}{0.85 ,0 ,0}

\newcommand{\DTNR}{DT$_\text{NR}$\xspace}
\newcommand{\DTR}{DT$_\text{R}$\xspace}
\newcommand{\DPNR}{${\text{DP}_\text{NR}}$\xspace}
\newcommand{\DPR}{{DP$_\text{R}$}\xspace}

\ifCLASSINFOpdf
\else
\fi

\usepackage{balance}

\begin{document}
%
\title{\LARGE Optical Visualization of Radiative Recombination at Partial Dislocations in GaAs}

\author{\IEEEauthorblockN{Todd~Karin$^1$, Xiayu~Linpeng$^1$, Ashish~K.~Rai$^2$, Arne~Ludwig$^2$, Andreas~D.~Wieck$^2$, Kai-Mei~C.~Fu$^{1,3}$} \\
\IEEEauthorblockA{$^1$Department of Physics, University of Washington, Seattle, Washington 98195, USA \\ 
$^2$Lehrstuhl f\"ur Angewandte Festk\"orperphysik, Ruhr-Universit\"at, D-44870 Bochum, Germany\\
$^3$Department of Electrical Engineering, University of Washington, Seattle, Washington 98195, USA
}}


%


\maketitle

\begin{abstract}
Individual dislocations in an ultra-pure GaAs epilayer are investigated with spatially and spectrally resolved photoluminescence imaging at 5~K. We find that some dislocations act as strong non-radiative recombination centers, while others are efficient radiative recombination centers. We characterize luminescence bands in GaAs due to dislocations, stacking faults, and pairs of stacking faults. These results indicate that low-temperature, spatially-resolved photoluminescence imaging can be a powerful tool for identifying luminescence bands of extended defects. This mapping could then be used to identify extended defects in other GaAs samples solely based on low-temperature photoluminescence spectra.

\end{abstract}


%
\IEEEpeerreviewmaketitle

\begin{figure*}
\centering
\includegraphics{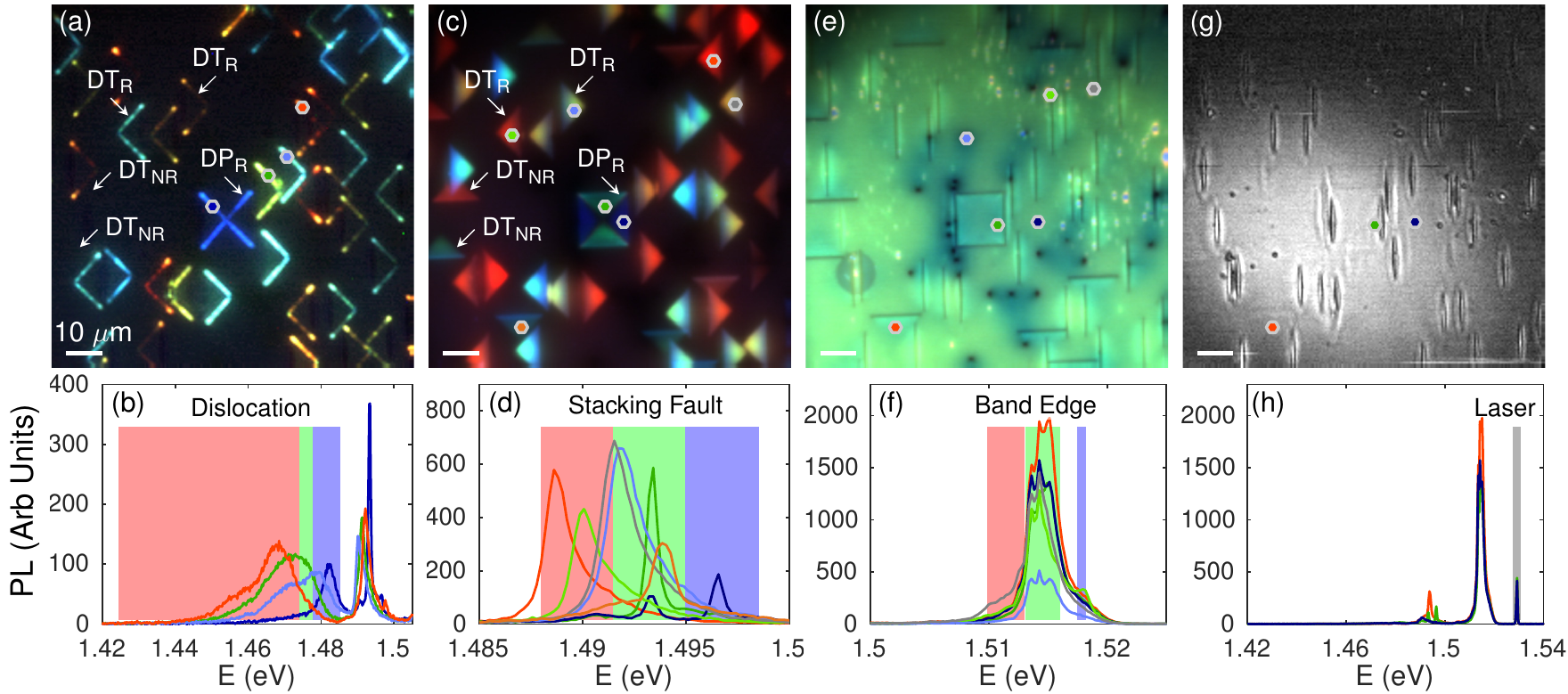}
\caption{ (a) Confocal scan of dislocations bordering stacking faults. The image is formed by coloring emission in different wavelength bands as red, blue or green, as depicted in b. Both radiative and non-radiative dislocations are observed. (b) Spectra of the sample at colored dots in a. PL is observed at the dislocations bordering the stacking faults. (c-d) Same as a-b, except with integration bands centered on the stacking fault bound exciton PL. (e-f) Same as a-b except with integration bands centered on the band edge PL. Dark lines are observed where the stacking faults intersect the sample surface. (f-h) Laser reflection image of sample. The vertical lines are oval defects.
Scale bar 10~\(\mu\)m. Excitation at 1.53~eV, $100~\mu$W, 1.9~K, excite and collect horizontal polarization.
}
\label{fig:confocal}
\end{figure*}

\section{Introduction}

Multijunction photovoltaics have been the leader in solar cell efficiency for over 20 years~\cite{NREL}. For high-efficiency multi-junction photovoltaics, the constituent semiconductor bandgaps must be precisely tuned to the solar spectrum and the resulting heterostructure must be grown with low dislocation densities. These requirements cannot be simultaneously satisfied using lattice-matched materials~\cite{Tibbits2014}, leading to a trade-off between more-optimal band gaps and higher dislocation densities due to lattice mismatch. Currently the highest efficiency cells avoid this issue either by wafer bonding~\cite{Tibbits2014}, or by using compositionally-graded buffer layers to minimize the presence of dislocations in specific regions~\cite{Geisz2006}. 
Since energy devices often contain dislocations that degrade device quality, it is advantageous to develop new tools for characterizing material quality.

In this work, we use a model system to investigate the radiative and non-radiative properties of individual dislocations and stacking faults. We use photoluminescence mapping to image 10~$\mu$m-scale extended defects in an ultra-pure GaAs epilayer, finding that different types of dislocations show markedly different luminescence behaviors. This non-destructive technique can be used to characterize individual dislocations in samples where the dislocation density is low compared with the optical resolution (\({\lesssim 10^7~\text{cm}^{-2}} \)). This work is in stark contrast to other studies of dislocations, which typically explore average material properties as a function of defect densities. Further, the knowledge gained from unambiguous studies of defects in ultra-pure samples can be used to characterize materials where single defects cannot be optically isolated.


\begin{figure}
\centering
\includegraphics{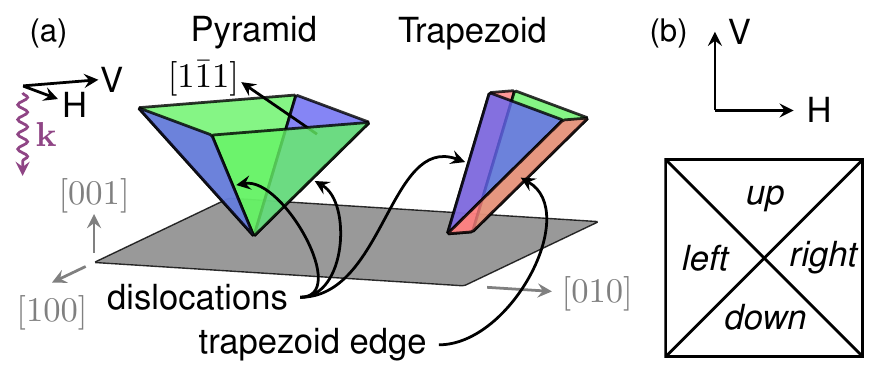}
\caption{(a) Two types of stacking fault structures are observed. Stacking fault pyramids consists of four stacking faults (colored planes) bounded by dislocations. Stacking fault trapezoids consist of a closely spaced ($\sim10~\text{nm}$) pair of intrinsic and extrinsic stacking faults, bounded by dislocations. H and V define the light polarizations and \(\mathbf k\) the wavevector of excitation light. (b) Experimental geometry with labeling scheme for stacking faults.}
\label{fig:sfdiagram}
\end{figure}

\section{Experiment}

\subsection{Sample preparation}

Stacking fault (SF) structures and associated dislocations are embedded in a 10-\(\mu\)m-thick [001] GaAs epilayer grown by molecular beam epitaxy, see Ref.~\cite{Karin2016arxiv} for details. The stacking fault defects nucleate at the substrate-epilayer interface during epitaxial growth~\cite{Kakibayashi1984,Chai1985}. Two kinds of defects are observed, see Fig.~\ref{fig:confocal}, and are depicted schematically in Fig.~\ref{fig:sfdiagram}: stacking fault pyramids and stacking fault trapezoids, which are each bordered by dislocations. We will refer to the dislocations surrounding the pyramid as \DPR and \DPNR and those surrounding the trapezoid as \DTR and \DTNR, where the subscript denotes whether radiative (R) or non-radiative (NR) transitions are observed.
Since we did not perform structural microscopy, we cannot confirm the identity of the dislocations observed. However, previous studies typically find stacking fault structures to be bordered by stair-rod partial dislocations~\cite{Wang1997,Abrahams1966,Wang1996,Finch1963}.

%

\subsection{Photoluminescence imaging}\label{sec:pl}

Figure~\ref{fig:confocal}(a) shows a low-temperature confocal scan of the sample in which photoluminescence (PL) is excited with an above-bandgap laser. The PL arises from excitons generated in the bulk that bind to extended defects and recombine radiatively.
 There are four main spectral bands of interest: PL from dislocations, PL from stacking faults, band edge PL, and laser reflection from the sample surface.

Strong photoluminescence is observed at 1.480~eV from the \DPR defect in Fig.~\ref{fig:confocal}(a), arising from excitons bound to the dislocation at the intersection of two SFs in a pyramid defect. We also observe luminescence due to excitons bound to the trapezoid edges, which is comprised of a dislocation-SF-dislocation structure as seen in Fig.~\ref{fig:sfdiagram}(a). PL at the trapezoid edges shows a large spread of \({\sim20~\text{meV}}\) in the center of the PL emission energy. We attribute this to the variable distance between the two edge dislocations in a trapezoid defect~\cite{Fung1997}, leading to a variable binding energy of excitons to the edge.
Considering that the typical distance between stacking faults in a trapezoid defect is \(\sim14\)~nm~\cite{Fung1997,Ayers2007} and the typical SF-bound exciton size is $>10~$nm~\cite{Karin2016arxiv}, the dislocations must play a role in the exciton binding potential at the trapezoid edge. In particular, strong non-radiative recombination at a dislocation would quench the PL.
Therefore, we will also refer to the luminescence originating from the trapezoid edges as due to excitons bound to dislocations.

Moreover, at the low experimental temperature (${\sim2~\text{K}}$) even a small excitonic potential of \({\sim 1~\text{meV}}\) is sufficient to bind excitons. Due to the non-isotropic strain fields surrounding a dislocation, we expect nearly all dislocations to create an attractive potential either to the electron or the hole, thus leading to a bound exciton~\cite{Emtage1967}. Therefore, if no radiative recombination is observed at a dislocation, we conclude that the dislocation is a strong non-radiative recombination center.

\begin{figure}
\centering
\includegraphics{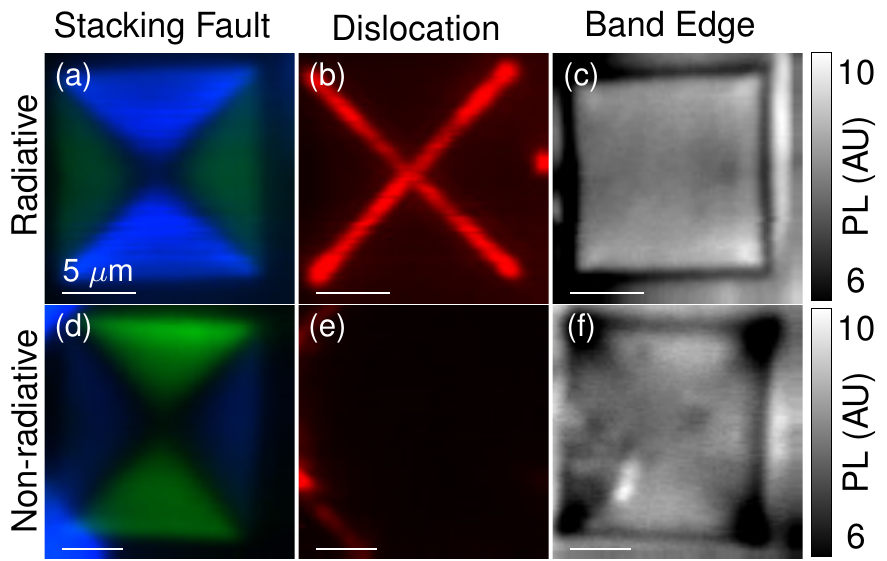}
\caption{(a-c) Confocal scan of the stacking fault pyramid visible in Fig.~\ref{fig:confocal} using integration bands green: 1.495--1.497~eV, blue: 1.492--1.494, red: 1.463--1.486~eV, grey: 1.511--1.520~eV. The stacking fault luminescence decreases where the stacking faults intersect the sample surface, but not where the stacking faults intersect one another. Excite at 1.52~eV, 1~mW, 1.47~K, excite H, collect H.
(d-f) Confocal scan of a SF pyramid that does not show dislocation luminescence in (e). Compared with (a), the stacking fault luminescence in (d) decreases more at the dislocations. The band edge luminescence is slightly quenched at the dislocations. Excite at 1.52~eV, 0.15~mW, 6.1~K, excite H, collect H.
}
\label{fig:radnonrad}
\end{figure}

Luminescence from stacking faults is shown in Fig.~\ref{fig:confocal}(c)-(d). In comparing Figs.~\ref{fig:confocal}(a) and \ref{fig:confocal}(c) we note that the \emph{up} and \emph{down} stacking faults show no PL from the dislocations at their edges, while the majority of the \emph{left} and \emph{right} stacking faults have strong PL from the adjacent dislocations [directions explained in Fig.~\ref{fig:sfdiagram}(b)]. This demonstrates that there are two types of partial dislocations, \DTR and \DTNR, with markedly different effects on charge carriers. These results are in contrast with the typical expectation that dislocations introduce strong non-radiative recombination centers~\cite{Yamaguchi1986,Lester1995}, but in line with related experiments where the type of dislocation changes the strength of non-radiative recombination observed~\cite{Petroff1980,Heinke1974,Shi2002,Bohm1979}.
Further, we observe pyramid defects where the dislocations are all \DPNR or all \DPR, see Fig.~\ref{fig:radnonrad}. In Figs.~\ref{fig:radnonrad}(d) and \ref{fig:radnonrad}(f), the stacking fault and band-edge PL appear quenched near the \DPNR defects.

The band-edge PL, Figs.~\ref{fig:confocal}(e),(f) quenches wherever the stacking faults intersect the surface. Surprisingly, this quenching seems unrelated to the surface oval defects, visible in Fig.~\ref{fig:confocal}(g)-(h), which only occur on the \emph{left} and \emph{right} stacking faults~\cite{Kasai1998}. Thus surface oval defects do not significantly affect the strength of non-radiative recombination occurring at the intersection of a stacking fault and the sample surface.

\begin{figure}
\centering
\includegraphics{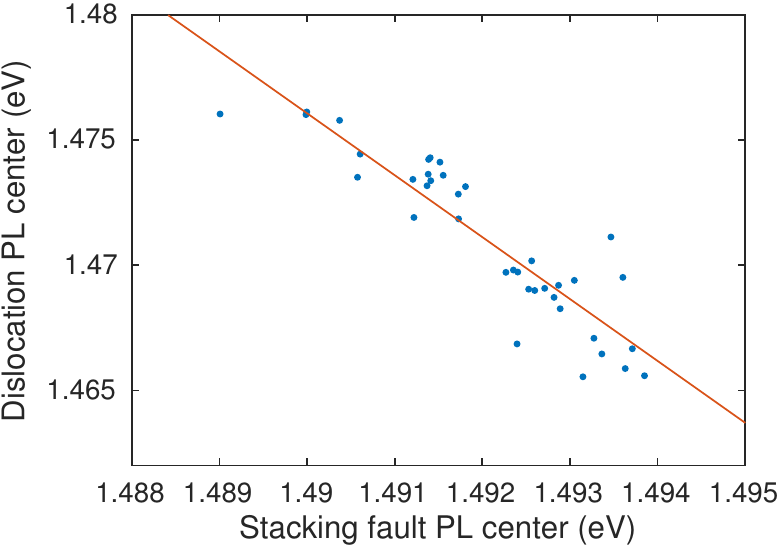}
\caption{Correlation of the dislocation and stacking fault PL center energies for the trapezoid defects in Fig.~\ref{fig:confocal}. The stacking fault PL center energy is inversely related to the corresponding dislocation PL center energy.}
\label{fig:corr}
\end{figure}

Lastly, we found the center-of-mass of the dislocation and stacking-fault PL for each trapezoid defect visible in Fig.~\ref{fig:confocal}, and performed a cross-correlation of these two emission energies (Fig.~\ref{fig:corr}). The PL center energy from the dislocation and the nearby stacking fault have a strong inverse correlation. This emission energy is related to the exciton binding energy, which is a function of the inter-stacking-fault distance in the trapezoid defect. Future work incorporating the full solution of the exciton effective-mass wavefunction could elucidate the physical mechanism behind this inverse correlation.



\section{Conclusion}

We use spatially and spectrally resolved exciton luminescence to investigate the properties of individual dislocations and stacking faults in GaAs. We find that some dislocations are efficient centers for radiative recombination while others have strong non-radiative recombination.
Future work combining structural microscopy and photoluminescence mapping can determine which dislocation varieties lead to radiative or non-radiative recombination.

This study was enabled by our unique sample, which contains a low density of 10-\(\mu\)m-scale extended defects embedded in a high-purity GaAs crystal. This enables us to determine the spectral characteristics of dislocations and stacking faults in an optical microscope.
The unambiguous identification of these defects may be useful for identifying stacking faults and dislocations in materials with high defect densities where individual defects cannot be optically isolated. For example, stacking fault densities could be estimated by comparing PL intensity to a reference sample. Lastly, we note that stacking faults and dislocations are present in a wide variety of zinc-blende and wurtzite materials, and that the same PL imaging techniques can be used to study other materials.

\section*{Acknowledgment}
 This material is based upon work supported by the National Science Foundation under Grant Number 1150647 and the National Science Foundation Graduate Research Fellowship under grant number DGE-1256082, and in part by the State of Washington through the University of Washington Clean Energy Institute.
A.K.R., A.L., and A.D.W. acknowledge partial support of Mercur Pr-2013-0001, DFG-TRR160, BMBF - Q.com-H  16KIS0109, and the DFH/UFA  CDFA-05-06.





%

\balance

\bibliographystyle{ieeetr}
\bibliography{toddkarin.bib}


\end{document}